\documentclass[aps,pra,letterpaper,twocolumn,showpacs,superscriptaddress,floatfix,longbibliograpghy]{revtex4-1}
\usepackage[english]{babel} \usepackage{amsmath} \usepackage{color}
\usepackage{epsfig} \usepackage{graphicx} \usepackage{bm}
\usepackage{mathtools}

\usepackage{times,amsmath,amssymb} \usepackage{amsfonts}
\usepackage{amssymb}

\usepackage[colorlinks,urlcolor=blue,bookmarks=false,hypertexnames=true]{hyperref}

 \newcommand{\ket}[1]{| #1 \rangle}
\newcommand{\scp}[2]{\langle #1 | #2 \rangle}
\newcommand{\braket}[3]{\langle #1 | #2 | #3 \rangle}

\definecolor{myred}{RGB}{168,5,14}
\definecolor{myblue}{RGB}{13,13,255}
\definecolor{mygreen}{RGB}{20,150,20}
\definecolor{editorcolor}{RGB}{168,5,14} 

\begin{document}

\title[GS energy universality of noninteracting fermionic systems]
{Ground-state-energy universality of noninteracting fermionic systems}

\author{Douglas F. C. A. Silva}\affiliation{Instituto de F\'isica,
  Universidade Federal da Bahia, Salvador--BA, 40170-115, Brazil}
  
\author{Massimo Ostilli}\affiliation{Instituto de F\'isica,
  Universidade Federal da Bahia, Salvador--BA, 40170-115, Brazil}

\author{Carlo Presilla}\affiliation{Dipartimento di Fisica, Sapienza
  Universit\`a di Roma, Piazzale A. Moro 2, Roma 00185, Italy}
\affiliation{Istituto Nazionale di Fisica Nucleare, Sezione di Roma 1,
  Roma 00185, Italy}

\date{\today}

\begin{abstract}
  When noninteracting fermions are confined in a $D$-dimensional
  region of volume $\mathrm{O}(L^D)$ and subjected to a continuous (or
  piecewise continuous) potential $V$ which decays sufficiently fast
  with distance, in the thermodynamic limit, the ground state energy
  of the system does not depend on {$V$}.
  Here, we 
  {discuss this theorem from several perspectives and derive a
    proof for radially symmetric potentials valid in $D$ dimensions}.
  We find that this universality property holds under a quite mild
  condition on $V$, {with or without bounded states,} and extends
  to thermal states. Moreover, it leads to an interesting analogy
  between Anderson's orthogonality catastrophe and first-order quantum
  phase transitions.
\end{abstract}

\maketitle

\section{Introduction and main result}
Noninteracting systems represent a crucial limit where analytic
solutions are possible and provide the zero{th}-order insight {into}
the understanding of more realistic models.  In particular,
noninteracting systems can be analyzed in the thermodynamic limit
(TDL), which provides a dramatically important link between
theoretical models (microscopic) and experiments (macroscopic). In
fact, the TDL is defined as the limit where both the number of
particles and the volume diverge while keeping their ratio constant.

The main result we are going to discuss in this work concerns the TDL
of the ground state (GS) energy of noninteracting fermionic systems
confined in some ``box''.  Let us consider a system of $N$
noninteracting fermions confined in a compact region of volume
proportional to $L^D$, where $L$ is a length and $D$ {is} the
dimensionality of the space considered. The corresponding one-particle
stationary Schr\"odinger equation must be solved with Dirichlet
boundary conditions, i.e., imposing that the eigenfunctions are zero
at the boundaries of the box (box with rigid walls). Let $E(N,L)$ be
the GS energy of this system.  Let us consider {then} another system
of $N$ noninteracting fermions confined in the same box {as in} the
previous system with the same boundary conditions but suppose that,
now, the fermions are subjected also to an external potential $V$.
Let us indicate by $\tilde{E}(N,L)$ the GS energy of this second
system.  The general result we want to discuss is that, for $L$ large,
the GS energies of these two systems differ at most by $o(L)$ terms
or, in other words, that in the TDL limit, where both the GS energies
(which are extensive observables) diverge as
$\mathrm{O}\left(L^D\right)$, their ratio tend to 1. More precisely,
for a spherically symmetric continuous (or more in general piecewise
continuous) potential $V(r)$, {where} $r$ is the radial distance, that
satisfies the {condition} {on the radial integral}
\begin{align}
  \label{main0}
  \int_0^{L} V(r)dr=\mathrm{O}\left(L^{1-\alpha}\right), \quad \alpha>0,
\end{align}
we have
\begin{align}
  \label{main2}
  \tilde{E}(N,L)=E(N,L)+\mathrm{O}\left(\rho L^{D-\alpha}\right), 
\end{align}
which, by using $E(N,L)=\mathrm{O}\left(L^D\right)$, provides
\begin{align}
  \label{main}
  \lim_{N\to\infty,L\to\infty,N/L^D=\rho}~\frac{\tilde{E}(N,L)}{E(N,L)}=1,
\end{align}
where $\rho=N/L^D$ stands for the particle density.  Alternatively,
Eq.~(\ref{main}) can be restated by saying that, in the TDL, the GS
energy per particle of the two systems is the same, and, in
particular, it depends on the boundary conditions but not on $V$
{(it is usually assumed that the boundary conditions do not
  matter in the TDL, which is true if $\alpha>D$, as proved rigorously
  in \cite{Courant-Hilbert}, as well as if $\alpha=D$, as proved
  rigorously in~\cite{Lieb}, however, for $\alpha<D$, no proof exists and boundary conditions might matter)}.
In Other words, at
zero temperature, given the boundary conditions and the
particle-density, we have the universality of the energy-density.
Moreover, we shall show that, in the case in which the integral in
Eq.~(\ref{main0}) is negative, the number of {eigenstates with
  negative energy} $N_0$ turns out to be nonextensive and scales as
\begin{align}
  \label{main1}
  N_0=\mathrm{O}\left( L^{D\left(1-\alpha/2\right)} \right),
\end{align}
whereas $N_0=0$ if $V\geq 0$, or at most $N_0=\mathrm{O}(1)$ when the
integral in Eq.~(\ref{main0}) is positive.

{ It is not hard to verify that Eq. (\ref{main2}) is consistent
  with the Lieb-Thirring inequalities in a box \cite{Lewin}, which
  provide rigorous bounds to the GS energy of a perturbed ideal gas
  via terms corresponding to the ``semiclassical-approximation''.
  However, the results in \cite{Lewin} do not cover the case $D=1$
  since the semiclassical-approximation there suffers from the
  presence of a Fermi-edge singularity~\cite{Fermi-edge}.  Our simple
  but general result is valid also for $D=1$ and unifies in a single
  group of scaling laws all the cases, including the cases with
    $\alpha\leq D$, i.e., the cases where the volume-{integral} of the potential
    diverges in the TDL.
  In particular,
  Eqs. (\ref{main2}) and (\ref{main1}) show the following: Whereas the
  energy shift caused by the potential
diverges for $L\to\infty$ (although non extensively)
  only when $\alpha<\alpha_E=D$, the threshold value for
  $N_0$ is $\alpha_0=2$, independently of $D$.  This implies that,
  for any given $D$, we can design experiments where the thermodynamic
  properties of the system with the potential are indistinguishable
  from those without the potential and, yet, unlike the latter, the
  former contains states that remain bounded in the TDL (although in a
  non extensive manner). In other words, we can accommodate an infinite
    number of electrons such that they are bounded around the minima of $V$,
    and still have a system
    that as a whole behaves as a perfect gas of fermions
    (and we can design such an experiment regardless of $D$, provided
    $\alpha<2$).
}

{As {the next section will make evident}, at the
  base of the above universality lie two facts: the absence of
  interaction and the exclusion principle. As is well known, the
  latter is the essential key for the stability of matter, as stressed
  long ago by Lieb in {his} monumental work~\cite{Lieb}. It is worth 
  {mentioning} however that, despite the obvious simpler nature of non
  interacting systems, their universality, in the sense of the
  Lieb-Thirring inequalities or in the sense of Eq. (\ref{main}),
  emerged only recently. In fact, the Lieb-Thirring inequalities in a
  box are the result of a formidable mathematical \textit{tour de
    force} that only experts in the spectral theory of Hilbert spaces
  are capable {of following}.  Here, we use a quite simpler strategy that
  allows to tackle also the $D=1$ case.}

The proof for $D=1$ {is based on a Pr\"ufer variables setting
  already used} in~\cite{Gebert}, which inspired us also to prove the
cases $D=2$ and $D=3$. As a technical note, we observe that
in~\cite{Gebert} and therein cited works, formulas for the difference
{in} the GS energies $\tilde{E}(N,L)-E(N,L)$ are expressed in terms of
integrals of the spectral shift function for one-particle
Schr\"odinger operators. The concept of the spectral shift function is
common in the spectral theory of Hilbert spaces \cite{Krein,Birman}
and in the foundations of Anderson' s orthogonality catastrophe
(AOC)~\cite{AOC} discussed in particular in
~\cite{GebertT,Gebert_AOC}, however, except for a Dirac-$\delta$
potential $V$, the formulas for the spectral shift function turn out
to be quite involved and hard to handle. In particular, it seems very
difficult to control the order of magnitude of the spectral shift
function with respect to $N$ and $L$ in the TDL, which is crucial for
proving Eq.~(\ref{main}).  An exception is the case $D=1$, where
in~\cite{Gebert} useful bounds for the spectral shift function are
explicitly worked out and used to express then the GS energy
difference in detail, which in particular implies
Eq.~(\ref{main}). This result of Ref.~\cite{Gebert} is, however,
obtained under the condition that $\int_0^L V(r)r^2 dr <\infty$, which
represents a much more stringent condition than Eq.~(\ref{main0}).

Here, we show how to deal with the cases $D=1,2$ and $3$ in a simpler
way that does not make use of the spectral shift function (for
convenience, the $D=2$ case is {discussed after} the $D=3$ case).  We start
demonstrating the general result (\ref{main}) by using an intuitive
heuristic argument that, at the same time, shows also why
Eq.~(\ref{main}) cannot hold for bosons. 
We then produce formal proofs and finally show how Eq.~(\ref{main})
suggests an interesting analogy between AOCs and first-order quantum
phase transitions~\cite{QPT}.
{Examples as well as counterexamples which shows that
Eq.~(\ref{main0}) constitutes a necessary and sufficient condition, are presented in Appendix.}

\section{Heuristic argument}
Let us consider a $D$-dimensional box of volume $L^D$ with hard walls
and $N$ noninteracting fermions subjected to a potential $V$ {with}
compact support over a region of volume $\ell^D$. Since the box has
hard walls, the wave function of each fermion must be zero outside the
box.  Moreover, the Pauli principle implies that the GS energy is the
sum of the first $N$ single-particle energies (accounting for their
possible degeneracies; note that, for simplicity, we assume spinless
fermions, however, if $D\geq 2$, the single-particle energies may have
degeneracies in the quantum numbers of the angular momentum).
Consider now a sequence of boxes of increasing volumes
$L^D_1,L_2^D,\ldots$ and corresponding number of fermions
$N_1,N_2,\ldots,$ such that the ratio $\rho=N_i/L^D_i$, $i=1,2,\dots$,
is kept constant. For {a sufficiently large index $i$} we will have
$\ell < L_i$ and eventually $\lim_{i\to\infty} \ell/L_i=0$.  On the
other hand, for $\ell \ll L_i$ the contribution of $V$ to the
single-particle energies will be $\mathrm{O}(\ell/L_i)$, at least for
those levels, corresponding to a sufficiently large radial quantum
number, for which the particles are, {on} average, homogeneously
distributed in the box. All these elements allow {us} to conclude that the
GS energy of the system in the presence of $V$ becomes closer and
closer to that of the pure box when $N$ and $L$ are larger and larger,
which is equivalent to Eq.~(\ref{main}).  It is easy to guess that the
condition that $V$ has a fixed compact support can be relaxed.  In
fact, in the formal derivation we shall show that the necessary and
sufficient condition is the quite mild one provided by
Eq.~(\ref{main0}).

The above heuristic argument {also} makes it evident that there is no
{analog} of Eq.~(\ref{main}) for noninteracting bosons. In fact,
their GS corresponds to a condensate in the lowest energy
single-particle state which 
implies that the particles distribute near the region where $V$ is
minimum.

\section{Proof of Eqs.~(\ref{main0}-\ref{main1}) by using Pr\"ufer
  variables}

{In Appendix, we consider explicit examples and show} how Eq.~(\ref{main})
realizes when $V$ is given by one or two Dirac-$\delta$ functions, and
it is plausible that similar results hold for any number of
Dirac-$\delta$ functions. Since any potential $V$ can be approximated
by a suitable sum of Dirac-$\delta$ functions, eventually in an
infinite number, in principle, one could attempt a general derivation
of Eq.~(\ref{main}) extending the above results. However, it seems
technically very hard to generalize the involved algebra to any number
of Dirac-$\delta$s and we prefer to resort to a different strategy
based on the Pr\"ufer variables~\cite{Prufer}, which is a common tool
within Sturm-Liouville's theory~\cite{Courant-Hilbert}.

We suppose that $V$ is piecewise continuous throughout the compact
region defining a box of side $L$ with rigid walls and assume the
validity of Eq.~(\ref{main0}).  Note that, these conditions, besides
implying the existence and regularity of the eigenfunctions of the
Schr\"odinger equation for any $L$~\cite{Courant-Hilbert}, imply also
that the number of eigenstates with negative energy $N_0$ (if any), is
finite for any $L$ and, as we shall see below, remains nonextensive in
the TDL, i.e., such eigenstates give no net contribution to the GS,
namely the general result (\ref{main1}).  Therefore, unless explicitly {stated otherwise}, in the following
we shall focus on {only} positive eigenvalues.

\subsection{$D=1$}
Consider the following two eigenvalue problems
\begin{align}
  \label{d1}
  -\frac{d^2 u}{d r^2}=k^2u, \qquad u(0)=u(L)=0,
\end{align}
\begin{align}
  \label{d1p}
  -\frac{d^2 u}{d r^2}+{\mathcal{V}}u=\tilde{k}^2u, \qquad u(0)=u(L)=0,
\end{align}
{where $\mathcal{V}=(2m/\hbar^2) V$,
  and} where $k$ and $\tilde{k}$ are real, i.e., we look for positive
eigenenergies $E=\hbar^2k^2/2m$ and $\tilde{E}=\hbar^2\tilde{k}^2/2m$,
respectively, Concerning Eq.~(\ref{d1}), the eigensolutions are given
by $u=A\sin(kr)$ and the boundary conditions imply the
``infinite-square-well'' values $k=n\pi/L$, with $n$ {being a} non null integer.
Concerning Eq.~(\ref{d1p}), following~\cite{Gebert}, we introduce
implicitly the Pr\"ufer variables, $\rho$ and $\theta$, by making the
following ansatz
\begin{align}
  \label{prufer}
  u(r)=\rho(r)\sin(\theta(r)),
\end{align}
\begin{align}
  \label{prufer2}
  u'(r)=\tilde{k}\rho(r)\cos(\theta(r)).
\end{align}
The physical idea of the above ansatz is that, when $V\to 0$, the
ansatz works with $\rho\to$constant, $\theta(r)\to kr$ and
$\tilde{k}\to n\pi/L$, while, for $V\neq 0$, $\rho$ is not constant,
$\theta(r)$ is not linear in $r$ and $\tilde{k}\neq n\pi/L$ but, in
general, $\delta=\theta(r)-\tilde{k}r$, which provides the so called
phase shift scattering, will {somehow be small, hence rendering} the
ansatz effective.  By making use of
Eqs.~(\ref{prufer})-(\ref{prufer2}) we see that Eq.~(\ref{d1p}) is
equivalent to the following system of first-order ODE
\begin{align}
  \label{d1pp}
  \rho'=\frac{1}{\tilde{k}}
  {\mathcal{V}}(r)\sin(\theta(r))\cos(\theta(r))\rho(r),
  \quad \rho(r)>0, 
\end{align}
\begin{align}
  \label{d1pp2}
  \theta'=\tilde{k}-\frac{1}{\tilde{k}}
  {\mathcal{V}}(r)\sin^2(\theta(r)),
  \quad \theta(0)=0, \quad \sin(\theta(L))=0.
\end{align}

{Although} the above first-order system is not useful for solving for $u$
(in fact, it is even non linear), it offers the best way to compare
$\tilde{k}$ and $k$ with each other, which is our main aim. Let us
integrate Eq.~(\ref{d1pp2}) between $r=0$ and $r=L$. By using the
boundary conditions we get $\theta(L)=n\pi=k_nL$ and obtain
\begin{align}
  \label{d1pp3}
  \tilde{k}_n=k_n+\frac{1}{L\tilde{k}_n}\int_0^L dr
  {\mathcal{V}}(r)\sin^2(\theta_n(r)),
  \qquad n=1,\ldots,N,
\end{align}
where we have inserted the dependencies on the integer index $n$ of
the eigenvalues running from 1 to the number of particles $N$.  Since
we are dealing with spinless fermions, for any such quantum numbers
$n$ we can allocate only one fermion.  Up to this point, we have
followed~\cite{Gebert}, where Eq.~(\ref{d1pp3}) is used to grasp the
$\mathrm{O}(1/L)$ corrections of the GS energy in a sophisticated
manner, where the $\mathrm{O}(1/L)$ terms are related to the decay
exponent of the AOC. Here, we follow a less demanding and simpler
strategy which is enough {for} our aims.  Let us rewrite
Eq.~(\ref{d1pp3}) as an equation for the eigenenergy $\tilde{E}_n$ as
a function of $E_n$. After {multiplying by $\tilde{k}_n$ and} squaring
the equation we get
\begin{align}
  \label{d1pp4}
  \tilde{E}_n^2-(E_n+2b)\tilde{E}_n+b^2=0,
\end{align}
where
\begin{align}
  \label{d1pp5}
  b={\frac{1}{L}}\int_0^L V(r)\sin^2(\theta_n(r)) dr.
\end{align}
For simplicity, in $b$ we have dropped a harmless dependence on
$n$. The important point to observe is that, under the condition
(\ref{main0}), we have $b=\mathrm{O}(1/L^{\alpha})$ for some
$\alpha>0$.  The roots of Eq.~(\ref{d1pp4}) are
\begin{align}
  \label{d1pp6}
  \tilde{E}_n=\frac{1}{2}E_n+b \pm \frac{1}{2}\sqrt{E_n^2+4bE_n}.
\end{align}
Note that, if $b<0$, for some $n$ the roots $\tilde{E}_n$ may {not be}
real.  When this occurs, it simply means that $\tilde{E}_n$ is
actually negative, against our initial assumption. However, by using
$E^2_n=\mathrm{O}(n^4/L^4)$ and $bE_n=\mathrm{O}(n^2/L^{2+\alpha})$,
we see that, if $b<0$, the number $N_0$ of eigenstates with negative
energy scales as $N_0=\mathrm{O}(L^{1-\alpha/2})$. This implies that,
as anticipated, the condition (\ref{main0}) negative implies
$N_0/N \to 0$, i.e., a non extensive $N_0$.  As a worst case example
where $\alpha=0$, consider a piecewise constant potential taking the
value $V(r)=V_0<0$ over a fractional portion of the box, $fL$, with
$0<f<1$, and $V(r)=0$ elsewhere.  In this case we obtain
$N_0\simeq\sqrt{2fm|V_0|L^2/\hbar^2}$, in agreement with known general
results~\cite{Davydov}.

Taking into account that the final target is to sum over
$n=1,\ldots,N$ in the TDL, with $N/L=\rho$ {being} constant, and by using
again $E^2_n=\mathrm{O}(n^4/L^4)$ and
$bE_n=\mathrm{O}(n^2/L^{2+\alpha})$, we see that in the square root of
Eq.~(\ref{d1pp6}) we can neglect the last term except for a number of
low-energy levels which scales as $N_0$, i.e., which can be
disregarded in the TDL.  Finally, choosing the root that is consistent
with the limit $V\to 0$, we arrive at
\begin{align}
  \label{d1pp7}
  \tilde{E}_n=E_n+2b + \ldots =
  E_n+\mathrm{O}\left(\frac{1}{L^\alpha}\right),
\end{align}
which implies
\begin{align}
  \label{d1pp8}
  \tilde{E}(N,L)=E(N,L)+\mathrm{O}(\rho L^{1-\alpha}),
\end{align}
and then Eq.~(\ref{main}).

Above, we have actually used an asymptotic expansion which might {not} sound
rigorous. However, we can use rigorous bounds as follows.  First,
observe that from the positive root of Eq.~(\ref{d1pp6}) we have
\begin{align}
  \label{Upper}
  \tilde{E}_n\leq E_n+2b.
\end{align}
On the other hand, if we now assume that $b\geq 0$, which in general
does {not} prevent the potential $V$ to take some negative values, we also
have
\begin{align}
  \label{Lower}
  \tilde{E}_n\geq E_n+b.
\end{align}
Equations (\ref{Upper}) and (\ref{Lower}) imply
$b\leq \tilde{E}_n-E_n\leq 2b$ leading again to Eq.~(\ref{d1pp8}).

\subsection{D=3}
In three dimensions, we express the Laplacian in spherical
coordinates {and} assume spherical boundary conditions and a spherical
symmetric potential $V=V(r)$. Given a spherical harmonic $Y_{l}^m$
with total angular momentum $l$ and magnetic momentum $m$, the
eigenvalue equations for the radial part of the wave function $R(r)$
written in terms of the function $u(r)=rR(r)$ read like
Eqs.~(\ref{d1}-\ref{d1p}) augmented with the ``centrifugal
contribution'':
\begin{align}
  \label{d3}
  -\frac{d^2 u}{d r^2}+\frac{l(l+1)u}{r^2}=k^2u,
  \quad u(0)=u(L)=0,
\end{align}
\begin{align}
  \label{d3p}
  -\frac{d^2 u}{d r^2}+{\mathcal{V}}u+\frac{l(l+1)u}{r^2}=\tilde{k}^2u,
  \quad u(0)=u(L)=0,
\end{align}
where, as before, $k$ and $\tilde{k}$ are real, i.e., we look for
positive eigenenergies $E=\hbar^2k^2/2m$ and
$\tilde{E}=\hbar^2 \tilde{k}^2/2m$, respectively. Concerning
Eq.~(\ref{d3}), the eigensolutions are given by the Bessel functions
$u=A j_l(kr)$ and the boundary conditions are satisfied by their
zeros, which are different from $k=n\pi/L$.  However, it is not
necessary to make use of the Bessel functions as we can workout the
$D=3$ case by using the previous setting of the Pr\"ufer variables
with suitable modifications as follows.  On comparing
Eqs.~(\ref{d3}-\ref{d3p}) with Eqs.~(\ref{d1}-\ref{d1p}), we see that
we cannot simply replace ${\mathcal{V}}$ with
${\mathcal{V}}+l(l+1)/r^2$ and repeat the $D=1$ argument since the
centrifugal term produces a diverging integral. Nevertheless, we can
integrate the analogous of Eq.~(\ref{d1pp2}) between some fixed
$\epsilon>0$ and $L$ and at the end send $\epsilon$ to zero suitably
to satisfy the lower boundary condition $\theta(0)=0$. We have to do
this {both for the cases} with and without $V$. Note that, now, the
Pr\"ufer variable $\theta$ refers to the case $V=0$ while for the case
$V\neq 0$ we shall use the symbol $\tilde{\theta}$. We have
\begin{align}
  \label{d3pp30}
  n\pi - \theta_n(\epsilon)
  =&\
     k_n(L-\epsilon)
     \nonumber \\
   &-\frac{1}{k_n}\int_\epsilon^L dr
     \frac{l(l+1)}{r^2}\sin^2(\theta_n(r)),
  \\
  \label{d3pp3}
  n\pi - \tilde{\theta}_n(\epsilon)
  =&\
     \tilde{k}_n(L-\epsilon)
     \nonumber \\ 
   &-\frac{1}{\tilde{k}_n}\int_\epsilon^L dr
     \left[{\mathcal{V}}(r)+\frac{l(l+1)}{r^2}\right]
     \sin^2(\tilde{\theta}_n(r)),
\end{align}
which gives
\begin{align}
  \label{d3pp3b}
  \tilde{k}_n =
  &\ k_n+
    \frac{1}{(L-\epsilon)\tilde{k}_n}\int_\epsilon^L dr
    \left[{\mathcal{V}}(r)+\frac{l(l+1)}{r^2}\right]
    \sin^2(\tilde{\theta}_n(r))
    \nonumber \\
  &-\frac{1}{(L-\epsilon)k_n}\int_\epsilon^L dr
    \frac{l(l+1)}{r^2}\sin^2(\theta_n(r))
    \nonumber \\
  &+ \frac{\theta_n(\epsilon)-\tilde{\theta}_n(\epsilon)}{L-\epsilon}.
\end{align}
We find it convenient to rewrite the above expression as follows
\begin{align}
  \label{d3pp3c}
  \tilde{k}_n =
  &\ k_n+
    \frac{1}{(L-\epsilon)\tilde{k}_n}\int_\epsilon^L dr
    {\mathcal{V}}(r)\sin^2(\tilde{\theta}_n(r))
    \nonumber \\
  &+ \frac{l(l+1)}{(L-\epsilon)\tilde{k}_n}\int_\epsilon^L dr\frac{1}{r^2}
    \left[\sin^2(\tilde{\theta}_n(r))-
    \frac{\tilde{k}_n}{k_n}\sin^2(\theta_n(r))\right]
    \nonumber \\
  &+ \frac{\theta_n(\epsilon)-\tilde{\theta}_n(\epsilon)}{L-\epsilon}.
\end{align}
Note that the dependence on $n$ inside the square {brackets} of the
integrand of the third term is weak and negligible with respect to the
dependencies {on} $n$ of $\tilde{k}_n$ and $k_n$.  We can now choose
$\epsilon$ as a suitable function of $L$ such that $\epsilon(L)\to 0$
for $L\to\infty$ and the third term becomes infinitesimal in the
TDL. For example, we can choose $\epsilon(L)=d/L^{1-\delta}$ with
$1>\delta>0$ and $d$ {being} constant. In this way the third term becomes
$\mathrm{O}(l^2/(\tilde{k}_nL^{\delta}))$.  Note also that, for
$L\to\infty$, the fourth term in the RHS of Eq.~(\ref{d3pp3c}) goes to
zero faster than the third one because both $\theta_n(\epsilon)$ and
$\tilde{\theta}_n(\epsilon)$ tend to $n\pi$. Therefore, we can repeat
the same steps as in Eqs.~(\ref{d1pp4}-\ref{d1pp6}) and reach the same
conclusions, with the only difference {being} that, now,
$b=\mathrm{O}(1/L^{\alpha})+\mathrm{O}(l^2/L^{\delta})$. The fact that
$\delta>0$ here is any arbitrary positive number, at most 1, just
means that, as soon as $l>0$, if $\alpha=0$, the centrifugal term
provides the leading correction to the GS energy, otherwise we choose
$\delta=\alpha$ so that the two corrections become equally dominant.
In either case, we are left with Eq.~(\ref{main2}).

\subsection{D=2}
In two dimensions, we express the Laplacian in polar coordinates 
{and} assume circular boundary conditions, and a symmetric potential
$V=V(r)$.  After separating the variables, the eigenvalue problems for
the radial part of the wave function become
\begin{align}
  \label{d2}
  -\frac{d^2 u}{d r^2}  +\frac{1}{r}\frac{d u}{d r}+\frac{m^2 u}{r^2}=k^2u,
  \quad u(0)=u(L)=0,
\end{align}
\begin{align}
  \label{d2p}
  -\frac{d^2 u}{d r^2}  +\frac{1}{r}\frac{d u}{d r}+\frac{m^2 u}{r^2}
  +{\mathcal{V}}u=\tilde{k}^2u, \quad u(0)=u(L)=0,
\end{align}
where the integer $m$ is the quantum number of the angular
momentum. Note that, unlike the cases $D=1$ (the infinite square-well
plus some potential) and $D=3$ (where $u/r$ represents the radial part
of the wave-function), for $D=2$ the boundary condition at $r=0$ does
not need to be zero (note that here $u(r)$ is the true radial part of
the wave function).  We could in fact choose any other value and reach
the same result. However, for simplicity and to keep conformity with
the previous cases, we choose $u(r)=0$.

Note that, unlike the $D=1$ and $D=3$ cases, Eqs.~(\ref{d2}-\ref{d2p})
contains also a term with the first derivative of $u$. We can however
still use the same definition of the Pr\"ufer variables according to
the ansatz (\ref{prufer}-\ref{prufer2}) which, once applied to
Eq.~(\ref{d2p}), generates the following first-order ODE (as for
$D=3$, the Pr\"ufer variables $\rho$ and $\theta$ refer to the case
$V=0$, while for the case $V\neq 0$ we shall use the symbols
$\tilde{\rho}$ and $\tilde{\theta}$)
\begin{align}
  \label{d2pp}
  \tilde{\rho}'
  =&\
     \left(\frac{1}{\tilde{k}}{\mathcal{V}}(r)\sin(\tilde{\theta}(r))
     \cos(\tilde{\theta}(r)) +
     \frac{m^2\sin(\tilde{\theta}(r))\cos(\tilde{\theta}(r))}{kr^2} \right.
     \nonumber\\
   &+ \left. \frac{\cos^2(\tilde{\theta}(r))}{r}\right)\tilde{\rho}(r),
\end{align}
\begin{align}
  \label{d2pp2}
  \tilde{\theta}'
  =&\
     \tilde{k}-\frac{1}{\tilde{k}}{\mathcal{V}}(r)
     \sin^2(\tilde{\theta}(r))
     -\frac{m^2\sin^2(\tilde{\theta}(r))}{kr^2}
     \nonumber \\
   &-\frac{\sin(\tilde{\theta}(r))\cos(\tilde{\theta}(r))}{r},
\end{align}
with the boundary conditions $\tilde{\rho}(r)>0$,
$\tilde{\theta}(0)=0$ and $\sin(\tilde{\theta}(L))=0$, and similarly
for the case with $V=0$ related to Eq.~(\ref{d2}).

By integrating Eq.~(\ref{d2pp2}) between $r=\epsilon$ and $r=L$ for
both the cases $V=0$ (symbols without the suffix $\tilde{\cdot}$) and
$V\neq 0$ (symbols with the suffix $\tilde{\cdot}$) and comparing, we
get
\begin{align}
  \label{d2pp3b}
  \tilde{k}_n
  =& k_n+
     \frac{1}{(L-\epsilon)\tilde{k}_n}\int_\epsilon^L dr
     {\mathcal{V}}(r)\sin^2(\tilde{\theta}_n(r))
     \nonumber \\
   &+ \frac{m^2}{(L-\epsilon)\tilde{k}_n}\int_\epsilon^L dr\frac{1}{r^2}
     \left[\sin^2(\tilde{\theta}_n(r))
     -\frac{\tilde{k}_n}{k_n}\sin^2(\theta_n(r))\right]
     \nonumber \\
   &+
     \frac{\theta_n(\epsilon)-\tilde{\theta}_n(\epsilon)}{L-\epsilon}
     \nonumber \\
   &+ \frac{1}{(L-\epsilon)}\int_\epsilon^L dr\frac{1}{r}
     \Big[ \sin(\tilde{\theta}_n(r))\cos(\tilde{\theta}_n(r))
     \nonumber\\
   &\qquad- \sin(\theta_n(r))\cos(\theta_n(r))\Big].
\end{align}
On choosing $\epsilon(L)=d/L^{1-\delta}$, with $1>\delta>0$ and $d$
{being} constant, and by observing that the latter term of the above equation
is $\mathrm{O}(\log(L)/L)$, which is subleading with respect to the
contribution coming from the centrifugal term proportional to $m^2$,
which is $\mathrm{O}(1/L^\delta)$, we reach the same conclusions {as in}
the case $D=3$~\cite{Note}.

{
  \section{AOC-QPT analogy}
  Our study is strictly connected to AOC.  In AOC, the GS overlap of
  two noninteracting fermionic systems defined with the same
  prescriptions discussed above, i.e., two systems of $N$ fermions
  confined in the same box of volume $L^D$, where the second system
  has also some potential $V$, 
  decays via a power law as follows
  \begin{align}
    \label{AOC}
    |\langle E(N,L)|\tilde{E}(N,L)\rangle|^2 \sim
    L^{-\gamma} = \left(N/\rho\right)^{-\gamma/D},
  \end{align}
  where $\ket{E(N,L)}$ and $\ket{\tilde{E}(N,L)}$ are the GSs of the
  finite size systems and $\gamma$ is a positive exponent that depends
  on the phase scattering shift.  Anderson originally derived
  Eq.~(\ref{AOC}) in Ref.~\cite{AOC} by using Hadamard's inequality
  for the determinant and a series of steps which were clarified and
  put on rigorous grounds only in recent
  years~\cite{Affleck1,Affleck2,GebertT}.  Equation~(\ref{AOC})
  implies that, in the TDL, the two GSs become orthogonal to each
  other while their energies, according to Eq.~(\ref{main}), become
  equal.

  It is interesting to consider the specific case of the
  Dirac-$\delta$ potential {(see Appendix for the case $D=1)$}. {Either} we can use
  the original informal bound of Ref. \cite{AOC} by
  evaluating the phase scattering shift from the energy shift
  via Fumi's theorem~\cite{Fumi}, or we
  can directly apply the following exact result just valid for a
  Dirac-$\delta$ potential~\cite{AOC2,Gebert_Delta} in $D=3$ with strength
  $\beta$:
  \begin{align}
    \label{AOCex}
    \gamma=\frac{\delta(\sqrt{\epsilon_F},\beta)^2}{\pi^2}, 
  \end{align}
  where $\epsilon_F$ is the Fermi energy (of the system without
  potential), and the phase shift is given by
  \begin{align}
    \label{phaseshift}
    \delta(\sqrt{\epsilon_F},\beta)=\left\{
    \begin{array}{l}
      \tan^{-1}\left(\sqrt{\frac{\hbar^{2} \epsilon_F}{2m}}
      \frac{1}{4\pi\beta}\right), ~\beta>0, \\
      \pi-\tan^{-1}\left(\sqrt{\frac{\hbar^{2} \epsilon_F}{2m}}
      \frac{1}{4\pi|\beta|}\right), ~\beta<0,  
    \end{array}
    \right.
  \end{align}
  or, in terms of the density
  \begin{align}
    \label{phaseshift1}
    \delta(\sqrt{\epsilon_F},\beta)=\left\{
    \begin{array}{l}
      \tan^{-1}\left(\frac{\hbar^{2}}{2m}\frac{\rho}{4\beta}\right),
      \quad \beta>0, \\
      \pi-\tan^{-1}\left(\frac{\hbar^{2}}{2m}\frac{\rho}{4|\beta|}\right),
      \quad \beta<0.  
    \end{array}
    \right.
  \end{align}
  Equation (\ref{phaseshift1}) shows that, as anticipated, the AOC {always}
  takes place, {regardless of how $\beta$ small is}.  But what is
  even more astonishing in Eq. (\ref{phaseshift1}) is that the phase shift
  is maximal in the limit of null strength, in other words,
  the resonance of the AOC
  occurs in the limit of null strength
  $\beta\to 0=\beta_r$, where
  $\delta(\sqrt{\epsilon_F},\beta)\to \pi/2$.

  This awkward property finds {an} interesting counterpart within
  first-order quantum phase transitions (QPTs). QPTs take place, at
  zero temperature in the TDL, when {one parameter (there can be more than one)} of the
  system, let's say $g$, assumes a critical values $g_\mathrm{c}$.  At
  the critical point $g=g_\mathrm{c}$, the GS energies of the two
  coexisting phases (the phase associated {with} the region
  $g>g_\mathrm{c}$ and that associated {with} the region $g<g_\mathrm{c}$)
  are equal, as prescribed by definition for a of QPT of first order,
  while the corresponding GSs become orthogonal, as {indicated} by the
  vanishing of the associated fidelity~\cite{Fidelity_Gu}.  A net
  example is the class of QPTs corresponding to a condensation in the
  space of states~\cite{QPT}, where the orthogonality of the GSs is
  granted by the separation of the Hilbert space in two complementary
  subspaces so that the scalar product in Eq.~(\ref{AOC}) vanishes (by
  construction it vanishes for any finite size but only the TDL of
  this construction is physical~\cite{QPT}).

  Equation
  (\ref{phaseshift1}) leads {us} to speculate that the resonance point
  $\beta_r=0$ corresponds somehow to a critical point. In this sense,
  the universality of the GS energies of noninteracting systems
  established by Eq. (\ref{main}) strongly supports this first-order
  QPT scenario: when the two phases ``meet'', they acquire the same
  energy, and at the same time becomes orthogonal and resonant to each
  other. Interestingly, a dynamical counterpart universality {was}
  recently confirmed in the framework of heavy impurities coupled to a
  Fermi sea, where the AOC is enriched by the presence of an active
  impurity, the impurity being part of the model with its own degree
  of freedom (as in the Anderson-Fano model)~\cite{Polarons}.  }

\section{Conclusions}
  In the TDL, the GS energy of a system of noninteracting fermions is
  universal, i.e., independent of an applied potential $V$, provided
  the latter decays sufficiently fast with distance. In the case of
  radially symmetric potentials, we have proved that, for dimensions
  $D=1,2,3$, the condition on $V$ is actually quite mild: the integral
  of $V(r)$ over the radial coordinate $r$ may even diverge as
  {$L^{1-\alpha}$} with any $\alpha>0$, which means that $V(r)$
  may decay with an arbitrary small power of $r$.

  The above result {was} obtained by representing the single-particle
  Schr\"odinger equations for $V=0$ and for $V\neq 0$ in terms of the
  Pr\"ufer variables and looking for rigorous inequalities between the
  respective single-particle eigenvalues $E_n$ and $\tilde{E}_n$.  For
  a finite size system of $N$ fermions in a box of volume $L^D$, we
  {found} $\tilde{E}_n = E_n + \mathrm{O}(1/L^\alpha)$, with $\alpha>0$
  representing the decay power of the potential $V$.  This result
  immediately provides the relation
  $\tilde{E}(N,L) = E(N,L) + {\mathrm{O}(\rho L^{D-\alpha})}$ for
  the GS energies of the two systems, and leads to the equality of the
  corresponding energies in the TDL, reached by sending $N$ and $L$ to
  infinity while {keeping} the density $\rho=N/L$
  constant. {Moreover, the same derivation allows us to establish
    that, in the case of non positive potentials, the number of
    bounded states scales as
    $N_0=\mathrm{O}( L^{D\left(1-\alpha/2\right)} )$.}

  It is clear that the TDL equality applies not only to the GS
  energies but also to the energies of an equilibrium thermal state,
  e.g., the canonical one obtained by summing the single-particle
  energies $E_n$ and $\tilde{E}_n$ with thermal weights given by the
  Fermi function.

  {Our study is strictly connected to AOC and definitively shows
    that, in any AOC,
    the GS energies of the two systems become equal in the TDL. This
    result thus adds in favor of an AOC - QPT analogy, where a power
    law decay of the GS overlap with certain critical
    exponents {({which} may be universal or not)} {is a common factor}. Furthermore, our analysis leads {us} to
    speculate that the resonant points of the AOC correspond somehow
    to the critical points of the QPT.

    The analogy between AOCs and QPTs is intriguing but raises several
    questions. Since an AOC is definitely not a QPT, what is the
    physical reason for this analogy? In {what} sense is an AOC
    analogous {to} but different from an actual QPT?  Is the limit of null
    strength always the resonant point in any AOC?  These and other
    interesting issues will {hopefully be} the subject of future
    work.}
 
  \begin{acknowledgments}
    We thank CNPq (grant 307622/2018-5) for
    funding. {This study was financed in part by the Coordena\c{c}\~{a}o de Aperfeiçoamento de Pessoal de N\'ivel Superior –
      Brasil (CAPES)
– Finance Code 001.}
    {We thank M. Lewin for pointing out to us Ref.~\cite{Lewin} and useful discussions}.
  \end{acknowledgments}

  \appendix*
\section{Illustrative examples and counterexamples}

\subsection{One Dirac-$\delta$ Confined  $(\alpha=1)$}
To illustrate the universality of the GS energy, we take the
straightforward example of a Dirac-$\delta$ potential in an infinite
square-well in $D=1$. Consider a particle of mass $m$ confined in the
interval $x\in (-L/2,L/2)$ via hard walls and potential
\begin{align}
  V(x)= -\beta\delta(x).
  \label{17}
\end{align}
The two eigenvalue problems we are interested in are then
\begin{align}
  \label{18a}
  -\frac{\hbar^{2}}{2m}\frac{d^{2}\psi}{dx^{2}}=E\psi(x),
  \quad \psi(\pm L/2)=0,
\end{align}
and
\begin{align}
  \label{18}
  -\frac{\hbar^{2}}{2m}\frac{d^{2}\psi}{dx^{2}}+V(x)\psi(x)=\tilde{E}\psi(x),
  \quad \psi(\pm L/2)=0.
\end{align}
The problem with $V=0$, Eq.~(\ref{18a}), has the well known
eigenvalues
\begin{align}
  E_{n} = \frac {{n}^ {2} \pi^{2} \hbar ^ {2}} {2mL^{2}},
  \qquad n=1,2,3,\dots.
  \label {001a}
\end{align}
Let us resolve the problem with $V\neq 0$.  Integrating \eqref{18}
around zero gives the discontinuity of the first derivative of the
wave function
\begin{align}
  \psi'(0^{+})-\psi'(0^{-})=-\frac{2m\beta}{\hbar^{2}}\psi(0)
  \label{23}.
\end{align}
Since $V$ is even, we can look for even and odd eigenfunctions.

Let us first focus on positive eigenvalues and let $k\in \mathbb{R}$.
For the even eigenfunctions we have
\begin{align}
  \psi(x)=
  \begin{cases}
    A\sin(kx)+B\cos(kx), &-L/2< x < 0, 
    \\
    -A\sin(kx)+B\cos(kx), \quad &0<x < L/2, 
  \end{cases}
\end{align}
where $A$ and $B$ are suitable constants and
$\tilde{E}=\hbar^{2}k^2/(2m)$.  On using \eqref{23}, the boundary
conditions $\psi(x=-L/2)=\psi(x=L/2)=0$, and defining $z\equiv kL/2$,
we obtain
\begin{align}
  \tan z=\frac{2\hbar^{2}}{m\beta L}z.
  \label{32}
\end{align}
In the limit $ L \rightarrow \infty $, Eq.~\eqref {32} has the
solutions $z=n\pi $, independent of $\beta $.  More precisely, for $L$
large but finite, we get the energy spectrum
\begin{align}
  \tilde{E}_{n} = \frac {(2n)^{2} \pi^{2} \hbar^{2}}
  {2mL^{2}} +\mathrm{O}\left(\frac{\hbar^{4} n^2}{m^2\beta L^3}\right),
  \qquad {n=1,2,3,\dots .}
  \label {001}
\end{align}
Note that in the above expressions we assumed $\beta\neq 0$, on the
other hand if $\beta=0$ we get exactly $z=(2n+1)\pi /2$ and we recover
the eigenenergies (\ref{001a}) with odd quantum number.  For the odd
eigenfunctions we have $\psi(x=0)=0$ so that, in this case, no
discontinuity of $\psi'$ applies and the solutions coincide with those
of the pure square-well potential:
\begin{align}
  \psi(x) = A\sin(kx),
\end{align}
\begin{align}
  \tilde{E}_{n} = \frac{(2n)^{2} \pi ^ {2} \hbar ^ {2}}{2mL^{2}},
  \qquad{n=1,2,3,\dots .}
  \label{002}
\end{align}

Let us consider now negative eigenvalues and let $K\in \mathbb{R}$.
The even eigenfunctions take the form
\begin{align}
  \psi(x) =
  \begin{cases}
    Ae^{Kx}+Be^{-Kx}, &\qquad -L/2<x<0, 
    \\
    Be^{Kx}+Ae^{-Kx}, &\qquad 0<x<L/2, 
  \end{cases}
\end{align}
where $A$ and $B$ are suitable constants and
$\tilde{E}=-\hbar^{2}K^2/(2m)$.  By imposing the continuity of $\psi$
in $x=0$, the boundary conditions in $x= \pm L/2$, the discontinuity
of $\psi'$, Eq.~(\ref{23}), and introducing $Z\equiv KL/2$, we achieve
the following transcendental equation for $Z$
\begin{align}
  \tanh Z =\frac{2\hbar^{2}Z}{m\beta L}, \label{144}
\end{align}
Eq.~\eqref {144} admits one single non trivial solution which exists
under the condition $b {\equiv (2 \hbar^{2})/(m \beta L)}<1 $, which,
in turn, is certainly satisfied for $L\rightarrow \infty$ where we get
$Z\to \pm [3(1-b)]^{1/2}$ and, therefore,
\begin{align}
  \tilde{E}_0=-\frac{6\hbar^{2}}{m L^2}+
  \mathrm{O}\left(\frac{\hbar^{4}}{m^2\beta L^3}\right).
  \label{neg}
\end{align}
Finally, concerning the odd eigenfunctions, it is easy to see that
they do not exist.

Consider now {$N$} noninteracting fermions in the infinite
square-well. The GS energy $E(N,L)$ is given by
\begin{align}
  E(N,L)
  &= \sum_{n=1}^{N}\frac{n^{2}\pi^{2}\hbar^{2}}{2mL^{2}}
    \nonumber\\
  &=
    { \frac{\pi^{2}\hbar^{2}}{2mL^{2}}
    \left( \frac{N^3}{3}+\frac{N^2}{2}+\frac{N}{6} \right).}
    \label{102a}
\end{align}
Similarly, in the square-well including the Dirac-$\delta$ potential,
by using Eqs.~(\ref{001}) and (\ref{002}) as well as Eq.~(\ref{neg})
(which contributes as a single eigenstate, but with the lowest energy,
to the GS), we obtain
\begin{align}
  &\tilde{E}(N,L)
    \nonumber \\ 
  &\quad =
    2\sum_{n=1}^{N/2}\left[\frac{(2n)^{2}\pi^{2}\hbar^{2}}{2mL^{2}}+
    \mathrm{O}\left(\frac{\hbar^{4}n^2}{m^2\beta L^{3}}\right)\right]
    + \mathrm{O}\left(\frac{\hbar^{2}N^2}{m L^{2}}\right)
    \nonumber \\
  &\quad =
    \left[ \frac{8 \pi^{2}\hbar^{2}}{2mL^{2}}
    + \mathrm{O}\left(\frac{\hbar^{4}}{m^2\beta L^{3}}\right) \right]
    \nonumber \\
  &\qquad \times
    \left( \frac{(N/2)^3}{3}+\frac{(N/2)^2}{2}
    +\frac{(N/2)}{6} \right) 
    +\mathrm{O}\left(\frac{\hbar^{2}N^2}{m L^{2}}\right) .
    \label{102}
\end{align}
On taking the ratio between \eqref{102} and \eqref{102a}, we get
\begin{align}
  \frac{\tilde{E}(N,L)}{E(N,L)}=
  { \frac{8 (N/2)^3}{N^{3}}
  +\mathrm{O}\left(\frac{\hbar^{2}}{m\beta L}\right)
  +\mathrm{O}\left(\frac{1}{N}\right). }
  \label{nplimit}
\end{align}
In the TDL, Eq.~(\ref{nplimit}) tends to 1, confirming the
universality expressed by Eq.~(\ref{main}).

\subsection{Two Dirac-$\delta$ Confined $(\alpha=1)$}
Now, let us consider a particle in the square-well with hard walls and
with two Dirac-$\delta$ perturbations symmetrically located at the
positions $\pm x_0$ in the $D=1$ square-well of length $L$ with
$L/2>x_0>0$.  The potential $V$ is then
\begin{align}
  V(x)= -\beta\delta(x-x_{0})-\beta\delta(x+x_{0}),
  \label{equation111}
\end{align}
and the boundary conditions $\psi(-L/2)=\psi(L/2)=0$. Let us focus on
positive eigenvalues and let $k\in \mathbb{R}$.  The eigenfunctions
can be written, in general, as
\begin{align}
  \psi(x)= \begin{cases}
    A_{1}e^{-ikx}+B_{1}e^{ikx}, \qquad &x\leq -x_{0},\\
    A_{2}e^{-ikx}+B_{2}e^{ikx},   &-x_{0}<x<x_{0},\\ 
    A_{3}e^{-ikx}+B_{3}e^{ikx},  &x\geq x_{0}. 
  \end{cases} \label{equation1111}
\end{align}
For even eigenfunctions, the above expression takes the form
\begin{align}
  \psi(x)= \begin{cases}
    A_{1}e^{-ikx}+B_{1}e^{ikx}, \qquad &x\leq -x_{0},\\
    A_{2}e^{-ikx}+A_{2}e^{ikx}, &-x_{0}<x<x_{0},\\ 
    B_{1}e^{-ikx}+A_{1}e^{ikx}, &x\geq x_{0}. 
  \end{cases} \label{equation1111e}
\end{align}
Applying the continuity of $\psi$ and the discontinuity of $\psi'$ in
$x=\pm x_{0}$, as well as the boundary conditions, we obtain the
following eigenvalue equation
\begin{align}\frac{\hbar^{2}}{m\beta L}=\cos^2\left(\frac{2z
  x_{0} }{L}\right)\frac{\tan z}{z}-\frac{\sin\left(\frac{2z
  x_{0}}{L}\right)
  \cos\left(\frac{2z x_{0}}{L}\right)}{z},
  \label{continuity4}
\end{align}
where $z=kL/2$.  For $ L \rightarrow \infty $, Eq.~\eqref{continuity4}
is reduced to
\begin{align} \frac {\tan z}{z} \rightarrow
  0 \label{continuity6}, \end {align} which has solution for
$ z=n\pi $.  More precisely, for $L$ large but finite, we get the
following energy spectrum
\begin{align}
  \tilde{E}_{n} &= \frac {(2n) ^ {2} \pi^{2} \hbar^{2}}
                  {2mL^{2}} +\mathrm{O}\left(\frac{\hbar^{4} n^2}{m^2\beta L^3}\right)
                  +\mathrm{O}\left(\frac{2x_0 \hbar^{2} n^2}{m L^3}\right),
                  \nonumber \\ &
                                 \qquad {n=1,2,3,\dots.}
                                 \label {001two}
\end{align}
Concerning the odd eigenfunctions, they take the form
\begin{align}
  \psi(x)= \begin{cases}
    A_{1}e^{-ikx}+B_{1}e^{ikx}, \qquad &x\leq -x_{0},\\
    2iB_{2}\sin(kx), &-x_{0}<x<x_{0},\\ 
    -A_{1}e^{ikx}-B_{1}e^{-ikx}, &x\geq x_{0}, 
  \end{cases} \label{equation1111o}
\end{align}
from which we obtain the following eigenvalue equation
\begin{align}
  &\frac{\hbar^{2}k}{2m\beta}\tan\left(\frac{kL}{2}\right)
    [\sin(kx_{0})+\cos^{2}(kx_{0})]
    \nonumber \\ &\qquad=
                   \tan\left(\frac{kL}{2}\right)[\cos(kx_{0})-\sin(kx_{0})].
                   \label{equationtg}\end{align}
                 The above equation admits the solution
                 $\tan (\frac{kL}{2}) = 0$, i.e. $kL = 2n \pi$,
                 {$n=1,2,3,\dots$}, which generates the same energy
                 spectrum {obtained at even quantum numbers for $V=0$}
                 \begin{align}
                   \tilde{E}_{n} = \frac {(2n)^{2}\pi^{2}\hbar^{2}} {2mL^{2}},
                   \qquad n=1,2,3,\dots .
                   \label{energia1}
                 \end{align}
                 Possibly, Eq.~\eqref {equationtg} has an extra
                 solution for
                 $ \frac {\hbar^{2} k}{2m \beta}[\sin (kx_ {0}) +
                 \cos^{2} (kx_ {0})] = [\cos (kx_ {0}) - \sin (kx_
                 {0})] $; however, this solution, if any, will not be
                 periodic and will not depend on $ L $.

                 We skip here the analysis for negative eigenvalues
                 since it can at most lead to the existence of two
                 solutions, which do not provide any net contribution
                 to the GS in the TDL.

                 By using Eqs.~\eqref{001two} and \eqref{energia1}, we
                 can proceed analogously to the previous section and
                 reach the following result for {the ratio between the
                   GS energies}
                 \begin{align}
                   \frac{\tilde{E}(N,L)}{E(N,L)}=
                   \frac{8(N/2)^{3}}{N^{3}} +
                   \mathrm{O}\left(\frac{\hbar^{2}}{m\beta L}\right) +
                   \mathrm{O}\left(\frac{x_0}{L}\right)
                   + \mathrm{O}\left(\frac{1}{N}\right).
                   \label{nplimittwo}
                 \end{align}

  \subsection{Counterexamples $(\alpha=0)$}
  It is easy to exhibit counterexamples to Eq.~(\ref{main}), all we
  need is to consider a potential $V$ whose integral scales as
  $\mathrm{O}(L)$.  To stay with the case just considered of an
  infinite-depth square-well in $D=1$, assume the same boundary
  conditions $\psi(-L/2)=\psi(L/2)=0$ and potential
  \begin{align}
    V(x)=
    \begin{cases}
      V_0, &\qquad x<0,
      \\
      0, &\qquad x>0,
    \end{cases}
           \label{ce1}
  \end{align}
  with, e.g., $V_0>0$. A detailed calculation of the single-particle
  eigenvalues can be performed as in the previous sections. Here,
  however, we just provide simple qualitative arguments valid in the
  limits of $V_0$ large and $V_0$ small.

  The number of eigenstates in a well of depth $V_0$ and width $L$
  with eigenenergies $0<E<V_0$, is
  $\mathrm{O}(\sqrt{m V_0 L^2 /\hbar^2})$ \cite{Davydov}, therefore,
  in the TDL at density $\rho=N/L$ constant we can certainly
  accommodate $N$ fermions in these states provided
  $\sqrt{m V_0/(\hbar^2\rho^2)} \gg 1$.  Moreover, if this condition
  is satisfied, the lowest $N$ states can be approximated by those of
  an infinite-depth well.  In the present case, this means that
  $\tilde{E}_n$, for $n=1,\dots,N$, are approximated by
  Eq.~(\ref{001a}) with the substitution $L\to L/2$. We conclude that
  in the TDL it must result $\tilde{E}(N,L)/E(N,L) \to 4$.

  Conversely, if $\sqrt{m V_0/(\hbar^2\rho^2)} \ll 1$, all eigenstates
  have eigenenergies $E> V_0>0$ so that we can estimate the effect of
  the potential $V$ by standard perturbation theory.  For the single
  particle energies we have
  \begin{align}
    \label{pert}
    \tilde{E}_n = E_n + \braket{n}{V}{n} = E_n + V_0/2,
  \end{align}
  where $\ket{n}$ are the eigenstates of the infinite-depth well and
  we used $\scp{n}{n}=1$ and $|\scp{x}{n}|^2$ symmetric in
  $[-L/2,L/2]$.  We immediately conclude that
  $\tilde{E}(N,L)/E(N,L) \to 1+3mV_0/(\pi\hbar\rho)^2$ in the TDL.

  As another remarkable counterexample, consider a periodic
  potential. Clearly, Eq.~(\ref{main0}) is not satisfied so that, in
  agreement with Bloch's theorem, the potential $V$ matters and
  Eq.~(\ref{main}) does not hold.

\subsection{A general toy model $(\alpha \geq 0)$}
Above, we have shown examples where $\alpha=1$ and counterexamples
with $\alpha=0$. In order to provide an example with a generic
$\alpha>0$, we should in principle consider a potential $V(x)$ whose
integral is weakly divergent in the TDL; for example
$V(x)=c/(|x_0|^{\alpha}+|x|^{\alpha})$.  We can however greatly
simplify this task by considering again the previous counterexample,
Eq.~(\ref{ce1}), where we now allow for a parametric dependence of
$V_0$ on $L$
\begin{align}
  V_0=v_0 L^{-\alpha},
\end{align}
where $\alpha\geq 0$ and $v_0>0$ are two constants. If $\alpha=0$ we
recover the previous counterexample, if instead $\alpha>0$, for $L$
sufficiently large we have $\sqrt{m V_0/(\hbar^2\rho^2)} \ll 1$, i.e.,
all eigenstates have eigenenergies $E> V_0>0$ and, applying
Eq.~(\ref{pert}), we get
\begin{align}
  \label{pert1}
  \tilde{E}_n = E_n + \braket{n}{V}{n} = E_n + \frac{v_0}{2L^\alpha},
\end{align}
which leads to Eq.~(\ref{main2}). The analysis of this toy model
suggests that Eq.~(\ref{main0}) may represent a necessary and
sufficient condition for Eq.~(\ref{main}) to hold.

  %


\begin{thebibliography}{99}%
    
    {
      
    \bibitem{Courant-Hilbert} R.~Courant and D.~Hilbert, ``Methods of
      Mathematical Physics'', Vol. I, (Wiley Classic Ed., 1989).
      
    \bibitem{Lieb} E.~H.~Lieb, ``The stability of matter'',
      Rev. Mod. Phys. \textbf{48}, 553 (1976).
     
      
    \bibitem{Lewin} R.~L.~Frank, M.~Lewin, E.~H.`Lieb, R.`Seiringer,
      ``A positive density analogue of the Lieb–Thirring inequality'',
      Duke Math. J. \textbf{162} (3), 435-495 (2013).

    \bibitem{Fermi-edge} T.~Ogawa, A.~Furusaki, and N.~Nagaosa,
      ``Fermi-edge singularity in one-dimensional systems'',
      Phys. Rev. Lett. \textbf{68}, 3638 (1992).
  

    \bibitem{Gebert} M.~Gebert, ``Finite-size Energy of
      Non-interacting Fermi Gases'', Math. Phys. Anal. Geom.
      \textbf{18}, 27 (2015).  }
  
  \bibitem{Krein} M.~G.~Krein, ``On a trace formula in perturbation
    theory'', Mat. Sb. \textbf{33} (75), 597-626. (Russian) (1953).
  
  \bibitem{Birman} M.~Sh.~Birman and D.~R.~Yafaev, ``The spectral
    shift function. The work of M. G. Kreln and its further
    development'', Algebra i Analiz. \textbf{4}, 1 (1992);
    English transl.: St. Petersburg Math. J. \textbf{4}, 833 (1993).
    
    {
    \bibitem{AOC} P.~W.~Anderson, ``Infrared catastrophe in Fermi gases
      with local scattering potentials'', Phys. Rev. Lett. \textbf{18},
      1049 (1967).  }
    
  \bibitem{GebertT} M.~Gebert, Ph.D. Thesis, ``Spectral and
    Eigenfunction Correlations of Finite-Volume Schrödinger
    Operators'' (2015).

  \bibitem{Gebert_AOC} M.~Gebert, ~K\"{u}ttler, P.~M\"{u}ller,
    ``Anderson's orthogonality catastrophe'',
    Commun. Math. Phys. \textbf{329}, 979 (2014).

  \bibitem{QPT} M.~Ostilli and C.~Presilla, ``First-order quantum
    phase transitions as condensations in the space of states '',
    J. Phys. A: Math. Theor. \textbf{54} 055005 (2021).

  \bibitem{Prufer} H.~Pr\"ufer, “Neue Herleitung der
    Sturm-Liouvilleschen Reihenentwicklung stetiger Funktionen”,
    Math. Ann., \textbf{95}, 499-518 (1926).
  
  \bibitem{Note} Rigorously speaking, unlike the case $D=3$, we see
    that, if $\alpha=0$, for $m=0$ the latter term of
    Eq.~(\ref{d2pp3b}), $\mathrm{O}(\log(L)/L)$, becomes dominant with
    respect to the second term $\mathrm{O}(1/L)$. However, informally,
    the behaviors $\mathrm{O}(\log(L)/L)$ and $\mathrm{O}(1/L)$ are to
    be considered equivalent and both associated with a power law with
    exponent $\alpha=0$, so that Eq.~(\ref{main2}) holds.

  \bibitem{Affleck1} I.~Affleck, Nuc. Phys. B \textbf{58}, 35
    (1997).
    
  \bibitem{Affleck2} A.~M.~Zagoskin and I.~Affleck, J. Phys. A \textbf{30},
    5743–5765 (1997).
  
{
\bibitem{Fumi} F.~G.~Fumi, ``Cxvi. Vacancies in Monovalent Metals'',
  The London, Edinburgh, and Dublin Philosophical Magazine and Journal
  of Science, \textbf{46}, 1007 (1955).  }
  
  \bibitem{AOC2} P.~W.~Anderson, ``Ground state of a magnetic impurity
    in a metal'', Pys. Rev. \textbf{164}, 352 (1967).

    {
    \bibitem{Gebert_Delta} M.~Gebert, ``The asymptotics of an
      eigenfunction-correlation determinant for Dirac-$\delta$
      perturbations'', J. Math. Phys. \textbf{56}, 072110 (2015). }
          
  
  \bibitem{Fidelity_Gu} S.-J.~Gu, Int. J. Mod. Phys. B \textbf{24},
    4371 (2010).

    {
    \bibitem{Polarons} R.~Schmidt \textit{et al.},
      Rep. Prog. Phys. \textbf{81} 024401 (2018).  }

  \bibitem{Davydov} A.~S.~Davydov, ``Quantum Mechanics'', 2nd ed.
    (Pergamon Press, Oxford, UK, 1985).

  \end{thebibliography}
\end{document}